\documentclass[showpacs,showkeys,preprintnumbers,amsmath,amssymb,nofootinbib,aps]{revtex4}

\usepackage{graphicx}
\usepackage{dcolumn}
\usepackage{bm}
\begin{document}

\title{A laser gyroscope system to detect the Gravito-Magnetic effect on Earth}

\author{A. Di Virgilio}
\email{angela.divirgilio@pi.infn.it}
\affiliation{
INFN Sez. di Pisa, Pisa, Italy}

\author{K. U. Schreiber and A. ˆ' Gebauer}
\email{schreiber@fs.wettzell.de}
\affiliation{
Technische Universitaet Muenchen,
Forschungseinrichtung Satellitengeodaesie \\
Fundamentalstation Wettzell, 93444 Bad K\"otzting, Germany}
\author{J-P. R. Wells}
\email{jon-paul.wells@canterbury.ac.nz}
\affiliation{
Department of Physics and Astronomy, University of Canterbury, PB4800,
Christchurch 8020, New Zealand}
\author{A. Tartaglia}
\email{angelo.tartaglia@polito.it}
\affiliation{
Polit. of Torino and INFN, Torino, Italy}
\author{J. Belfi and N. Beverini}
\email{belfi@df.unipi.it, beverini@df.unipi.it}
\affiliation{Univ. of Pisa and CNISM, Pisa, Italy}
\author{A.Ortolan}
\email{antonello.ortolan@lnl.infn.it}
\affiliation{Laboratori Nazionali di Legnaro, INFN Legnaro (Padova), Italy}


\begin{abstract}
Large scale square ring laser gyros with a length of four meters on each side are
approaching a sensitivity of $1 \times 10^{-11} \mbox{rad}/\mbox{s}/\sqrt{\mbox{Hz}}$. This is about
the regime required to measure the gravito-magnetic effect (Lense
Thirring) of the Earth. For an ensemble of linearly independent gyros each measurement signal depends upon
the orientation of each single axis gyro with respect to the rotational axis of the Earth. Therefore at
least $3$ gyros are necessary to reconstruct the complete angular orientation of the apparatus.
In general, the setup consists of several laser gyroscopes (we would prefer more than $3$ for sufficient
redundancy), rigidly referenced to each other. Adding more gyros for one plane of observation
provides a cross-check against intra-system biases and furthermore has the advantage of
improving the signal to noise ratio by the square root of the number of gyros.
 In this paper we analyze a system of
two pairs of identical gyros (twins) with a slightly different orientation
with respect to the Earth axis. The twin gyro configuration has several
interesting properties. The relative angle can be controlled and provides a useful null
measurement. A quadruple twin system could reach a $1\%$
sensitivity after $3.2$ years of data, provided each square ring has $6$ $m$ length on a side,
the system is shot noise limited and there is no source for 1/f- noise.
\end{abstract}

%
\maketitle
\section{\label{sec:a} Introduction}
General relativity shows that the gravitational field of a rotating body is
different to that of a non-rotating body of equal mass and shape. The
measurement of such a gravito-magnetic (gm) effect is extremely difficult \cite{ciufolini, GPB},
since it is $10^9$ times weaker than the Sagnac signal generated by the rotation of the Earth.
Ring laser gyroscopes have the potential to measure absolute rotation rates. Therefore they are good
candidates for an experimental test of the gravito-magnetic
effect. Typical navigational ring laser gyros in aircrafts provide a sensor resolution around
$5 \times 10^{-7}\mbox{rad}/\mbox{s} /\sqrt{\mbox{Hz}}$ and the drift that can be experienced is as low
as $0.0001 \mbox{deg}/\mbox{h}$. These are excellent parameters for a gyroscope, but it is by
far not enough for either a precise monitoring of Earth rotation or the frame dragging effect.
The ring laser equation~\cite{geoff} relates the frequency splitting $\delta f$ of the two
counter propagating optical beams inside the interferometer with the rate of rotation $\Omega$ as:
\begin{equation}
\delta f = \frac{4 A}{\lambda P} {\bf n} \cdot {\bf \Omega} ,               \label {eq3.1}
\end{equation}
The expression $\frac{4 A}{\lambda P}$, describes the scale factor of the instrument and depends on the area (A)
circumscribed by the two laser beams, the corresponding perimeter (P) of the contour and the
wavelength ($\lambda$) acting as a "Vernier scale" in this optical device, while $n$ is the normal vector
on the optical plane of the instrument. As one can see from equation~\ref{eq3.1}
upscaling the size of the instrument is a good way of improving the sensitivity of the apparatus.
The large ring laser ``G'' \cite{ulli09} at the Geodetic Observatory in Wettzell (Bavaria, Germany)
has a square contour with an area of 16 m$^{2}$ and a corresponding perimeter of 16 m. It comprises a HeNe
cw monomode laser with a wavelength of $\lambda = 0.6328 \mu$m and is placed on a very stable granite
monument in a laboratory approximately 6 m below the Earth surface. Because of it's size and
substantial stabilisation of all relevant operational parameters, a performance level of better than
$1.26 \times 10^{-11} \mbox{rad}/\mbox{s}$ is now routinely obtained, which can resolve rotations of 1 pico-rad/s at
about 1 hour of integration. This is about one order of magnitude short of the requirements for the
detection of the gravito-magnetic effect. Very Long Baseline Interferometry (VLBI) measures the
Earth rotation rate with high accuracy, by obtaining precise directions to Quasars mapped out in the
International Celestial Reference Frame (ICRF). The superposition between the two independent
measurement systems, has the potential to isolate the Lense-Thirring term, because it is not present
in the VLBI measurement, while it is contained in the ring laser signals.

\section{\label{sec:b}   Gravitomagnetic effects}
The term gravitomagnetism is commonly used to designate the effects and
phenomena which happen in a gravitational field when the source of the field
is rotating\footnote{%
Gm effects in general can come from any mass current, but only the
rotational components cannot be eliminated by a coordinate transformation.}
and not when it is at rest in an inertial reference frame.   Actually
rotational effects may also be there not because of a gravitational field,
but simply because the observer is using a rotating reference frame:
typically this is the Sagnac effect. The latter effects are termed
kinematic and are not included in the gravitomagnetic ones. The
gravitational rotation effects are in general much much weaker than the so
called gravito-electric ones (whose zero order is Newton's law), so they are
treated by approximating the Einstein equations to the lowest order in the
angular momentum of the source. Strictly speaking only these linearized
contributions are called gravitomagnetic (gm) \cite{MRAT}.

In General Relativity the gravitational field is contained in the line
element of the space-time around a source of gravity. If the central mass is
steadily rotating the metric tensor has two symmetries: an axial symmetry in
space and an invariance in time. Usually the typical axially symmetric
stationary line element is approximated as%
\begin{equation}
ds^{2}\simeq \left( 1-2\frac{\mu }{r}\right) c^{2}dt^{2}-\left( 1+2\frac{\mu
}{r}\right) dr^{2}-r^{2}d\theta ^{2}-r^{2}\sin ^{2}\theta d\phi ^{2}+4\frac{j%
}{r^{2}}\sin ^{2}\theta d\tau \left( rd\phi \right)     \label{base}
\end{equation}%
Formula (\ref{base}) is valid for an inertial observer at rest with the
center of the rotating body. The space coordinates are polar Schwarzschild
coordinates. The dimensionless parameters (in the case of the surface of the
Earth) are

\begin{eqnarray}
\mu   &=&G\frac{M}{c^{2}}\simeq 4.4\times 10^{-3}\text{ m; \qquad \quad }%
\frac{\mu }{R}\sim 10^{-9}   \nonumber \\
j &=&G\frac{I}{c^{3}}\Omega \simeq 1.\,\allowbreak 75\times 10^{-2}\text{ m}%
^{2}\text{;\qquad }\frac{j}{R^{2}}\sim 10^{-15}   \nonumber
\end{eqnarray}%
and represent the reduced mass $\mu $ and angular momentum $j$ ($\Omega $ is
the angular velocity of the diurnal rotation of the planet). The numbers
give an idea of the comparative size of the gravitomagnetic effects,
accounted for by $j$, and the gravitoelectric ones expressed by the pure
mass terms. Typically $j$ produces a frame dragging, which was pointed out
in 1918 by Lense and Thirring \cite{thir} and goes under their names: in practice in
a strong field, such as the one in vicinity of a Kerr black hole, a locally
non-rotating observer would be seen as rotating by an inertial observer at
infinity .

We are interested in what can be measured in a terrestrial laboratory so we
need to convert (\ref{base}) to the reference frame of a local observer
located on Earth. This passage is made in two steps: a) from the initial
reference frame to an instantaneously co-moving inertial frame
(instantaneously translating at the peripheral speed of the laboratory about
the axis of the Earth); b) from the intermediate inertial frame to a
non-inertial rotating one (the scalar angular velocity is the one of the
Earth $\Omega $). The result is summarized in the new metric tensor:

\begin{eqnarray}
g_{00} &\cong &\left( 1-2\frac{\mu }{r}-\frac{\Omega ^{2}}{c^{2}}r^{2}\sin
^{2}\theta \right)     \nonumber \\
g_{rr} &\cong &-\left( 1+2\frac{\mu }{r}\right)     \nonumber \\
g_{\theta \theta } &=&-r^{2}   \label{metrica} \\
g_{\varphi \varphi } &\cong &-r^{2}\left( 1+\allowbreak 2r^{2}\frac{\Omega
^{2}}{c^{2}}\sin ^{2}\theta \right) \sin ^{2}\theta     \nonumber \\
g_{0\varphi } &\cong &\left( 2\frac{j}{r}-r^{2}\frac{\Omega }{c}-2\mu r\frac{%
\Omega }{c}\right) \sin ^{2}\theta     \nonumber
\end{eqnarray}

Another relevant parameter is now $R\Omega /c\sim 10^{-6}$.

\begin{figure}[th]
\includegraphics[width=271pt]{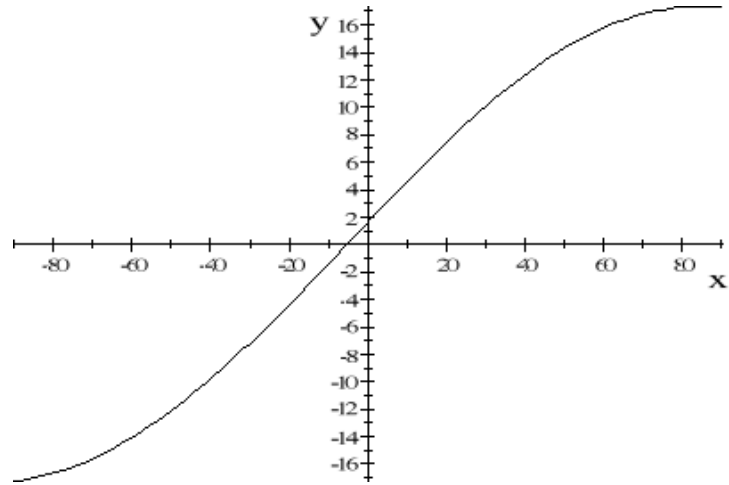}
\caption{Difference between the total signal and the pure Sagnac
contribution. Vertically beat frequencies are shown in units of $10^{-7}$
Hz. Horizontally angles in degrees are found. The zero corresponds to the $0$
Sagnac contribution.}
\label{fig1}
\end{figure}

\begin{figure}[th]
\includegraphics[width=271pt]{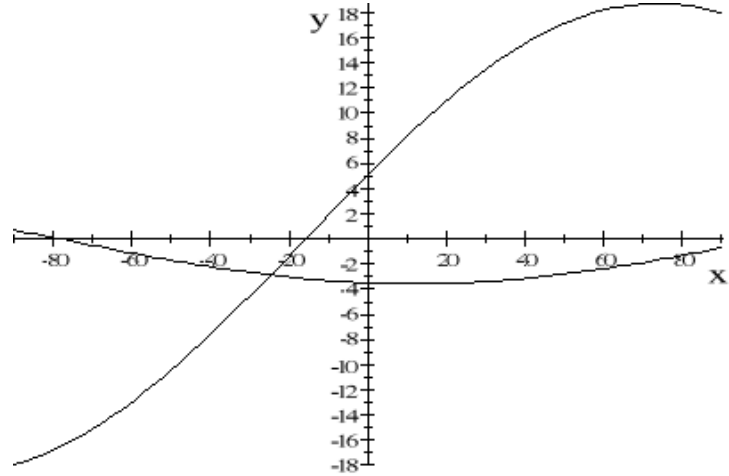}
\caption{Using the same variables as in fig. 1 the geodetic precession term
(upper) and the frame dragging term (lower) are shown.}
\label{fig2}
\end{figure}

In the field above two light rays moving along the same closed path in
opposite directions need different times of flight to complete the circuit.
We can compute the two times of flight by integrating the null line elements
corresponding to (\ref{metrica}) along the contour. A laser gyroscope is a
ring cavity with counter-propagating light beams. It converts the time of
flight difference between counter-propagating beams into a beat frequency of
the two travelling waves. The beat frequency depends on three separate
contributions: the Sagnac effect (since the reference is a non-inertial
one), the geodetic precession term (depending on the pure mass) and the
frame dragging term (depending on the angular momentum). Since the Sagnac
effective gm field is everywhere parallel to the axis of the Earth, whereas
the other two terms are not, the three contributions have a different
dependence on the orientation of the laser gyroscope. From simple geometrical
considerations it is easy to see that all effects become zero if the
instrument is contained in a meridian plane (all fluxes vanish). If $\hat{n}$
is a unit vector perpendicular to the plane of the beams, the analysis can
be restricted to the case in which $\hat{n}$ is contained in the meridian
plane. When the angle between $\hat{n}$ and the local radial direction
coincides with the local latitude, we expect the Sagnac contribution
to remain zero, because the plane of the laser gyroscope contains the direction of
the axis of the Earth, so this would in principle be the best configuration
for an experiment. However the instrument needs a bias, so this choice is
not viable. In general the beat frequency is
\begin{eqnarray}
\nu _{b} &\cong &4\frac{c}{\lambda P}\frac{S}{R}\mid \left( \frac{1}{c}%
R\Omega -\allowbreak 2\frac{j}{R^{2}}+4\frac{R\Omega }{c}\frac{\mu }{R}%
\right) \sin \left( \Theta +\psi \right) \cos \Theta     \label{frequenza} \\
&&-\left( \frac{1}{c}R\Omega +\allowbreak \frac{j}{R^{2}}+2\frac{R\Omega }{c}%
\frac{\mu }{R}\right) \cos \left( \Theta +\psi \right) \sin \Theta \mid
\nonumber
\end{eqnarray}%
where $\Theta $ is the co-latitude of the laboratory and $\psi $ is the tilt
angle in the meridian plane from the zero Sagnac configuration. $S$ is the
area of the closed path of light, $P$ its length; $\lambda $ is the
wavelength of light. Fig. \ref{fig1} shows the difference between the expected total signal and
the pure Sagnac term, evidencing the sum of the geodetic precession and the
Frame Dragging terms, as functions of the tilt angle in the meridian plane.
A colatitude $\Theta =47%
{{}^\circ}%
$ (corresponding to the Gran Sasso laboratories) has been used; $S=36$ m$%
^{2} $ and $P=24$ m. The two gm terms together become zero at an angle $\psi
_{p}\simeq -5.51%
{{}^\circ}%
$ .
Fig. \ref{fig2} displays separately the geodesic term (bigger) and the frame
dragging term (smaller).
We see that each term becomes zero at different positions: respectively at
$%
\psi _{g}\simeq -15.98%
{{}^\circ}%
$ and $\psi _{LT}\simeq -78.02%
{{}^\circ}%
$.
Finally one sees that the full signal becomes $0$ where the residual Sagnac signal
is compensated by the two other terms; this happens at the angle $\psi
_{0}\simeq -2.\,\allowbreak 7\times 10^{-10}$ rad. This position, as already
stated, is practically not accessible.

Of course the possibility of isolating the gm terms relies both on the
different angular dependence and the possibility to accurately model the
Sagnac effect. In practice we need an independent way to measure the angular
speed of the laboratory, $\Omega $. The latter is with respect to fixed
stars so that it is indeed the result of a superposition of the rotations
about the axis of the planet and around the Sun, and may change in time. The
independent measurement we need can be based on VLBI and requires an
accuracy of at least $1$ part in $10^{9}$. Since the angles also play an
important role in determining the contribution of each component, we
need to know the orientation of the rotation axis with an accuracy better
than $10^{-9}$ rad. The accurate knowledge of $\Omega $ and $\psi $ would
allow for the subtraction of the Sagnac bias due to Earth rotation, thus leaving the proper gm
effects, that would be distinguished from one another by measurements made
at the angles $\psi _{g}$ and $\psi _{LT}$.

\section{\label{sec:c} Sensor properties}

A closer look at equation~\ref{eq3.1} reveals that there are three basic effects one has to carefully
account for. These are:
\begin{itemize}
\item scale factor stability ($4 A/\lambda P$)
\item orientation of the gyroscope with respect to the instantaneous axis of rotation of the Earth
\item instantaneous rate of rotation of the Earth -- Length of Day: (LoD)
\end{itemize}
The scale factor for all practical purposes has to be held constant to much better than 1 part in
$10^{10}$. Otherwise the frame-dragging parameter cannot be determined unambiguously. For G, the
base of the gyroscope has been manufactured from Zerodur, a glass ceramic with a thermal expansion
coefficient of $\alpha < 5 \times 10^{-9}/^{o}C$. Furthermore the instrument is located in a thermally
insulated and sealed environment with typical temperature variations of less than 5 mK per day. However, because
the underground laboratory is only at a depth of 5~m, there is still a peak to peak temperature
variation of about 1 degree per year, accounting for the change of seasons as the Earth circles around
the sun. Changes in the atmospheric pressure also affect the dimensions of the ring laser structure
by changing the compression of the Zerodur block
and cannot be neglected. Hence G is kept in a pressure stabilized enclosure. A feedback system
based on the determination the current value of the optical frequency of the lasing mode of one sense of propagation
allows for actively control of the pressure inside the steel vessel such that an overall geometric scale
factor stability of better than $10^{-10}$ is routinely obtained.  At the same time the design of the
instrument is made as symmetric as possible. So changes in area and perimeter are compensated
with a corresponding change in wavelength as long as no shear forces are present and the longitudinal
mode index stays the same. Whether this is good enough for the requirement envisaged remains to be shown.
Our middle size ring laser G-Pisa ( $1.35$ $m$ side) \cite{amaldi}, at the moment installed in the central area of the Virgo interferometer for gravitational wave research, is testing a perimeter control scheme.
Since we actively control the frequency of the
lasing mode, this stabilization system compensates small uniformly acting temperature effects as well.
Temperature gradients however are not compensated and need to be avoided.
Increasing the scale factor works well for moderate size ring lasers.
Systems larger than 100 m$^{2}$ of area suffer strongly from mechanical instabilities \citep{Benni, geoff1}
and have shown inferior stability, which is not compensated by the increased sensitivity.
At this point in time the optimum size for the ring laser structure is not yet known.
Active closed loop perimeter stabilization is expected to improve this situation substantially,
but this is work in progress.
Purely instrumental issues such as backscatter reduction, null-shift stability by reducing the variation
in laser mode pulling and pushing inside the ring laser cavity certainly remain to be the
most important effects to control. A wider discussion of these effects is given in~\citep{geoff, ulli09,bob_ug2} and \citep{rlg-noise}.

It has been shown in the theoretical section that the orientation of the ring laser is critical with
respect to measurement of the frame dragging parameter. The inner product in equation~\ref{eq3.1}
between the normal vector on the gyroscope and the instantaneous Earth rotation vector determines the
projection of the measurement quantity onto the apparatus. This projection needs to be known to an
accuracy of better than 1~nano-rad. In the presence of geophysical signals such as solid Earth tides
\citep{tides} and diurnal polar motion \citep{polmot}, ocean loading and atmospheric loading, it is
necessary to keep track of all the changes in ring laser orientation. Therefore it will require a
full triad system of linear independently orientated ring laser gyroscopes to obtain variations in
the ring laser orientation at the required level of accuracy. We expect that the misalignment between
each ring laser subsystem with respect to orthogonality has to be established interferometrically.

Earth rotation is not constant at the level required for the detection of the Lense-Thirring effect.
Therefore the changes caused by mass redistribution over the Earth and momentum exchange between
the lithosphere, atmosphere and hydrosphere have to be monitored as well. Apart from taking these
quantities as ``high'' frequency components superimposed on the constant frame dragging parameter,
the variation in LoD can be taken from the time-series of geodetic
Very Long Baseline Interferometry (VLBI) from the International Earth Rotation and Reference System
Service (IERS: {\it http://www.iers.org/}) independently with high precision. The VLBI technique is
currently almost at the required measurement resolution. Further improvements can be expected in the
near future since the VLBI 2010 design specifications of the International VLBI Service (IVS) are in
the process to be implemented.

\section{\label{sec:d}Current Ring Laser Results}

\begin{figure}[ht]
\centering
\includegraphics[height=5.4cm]{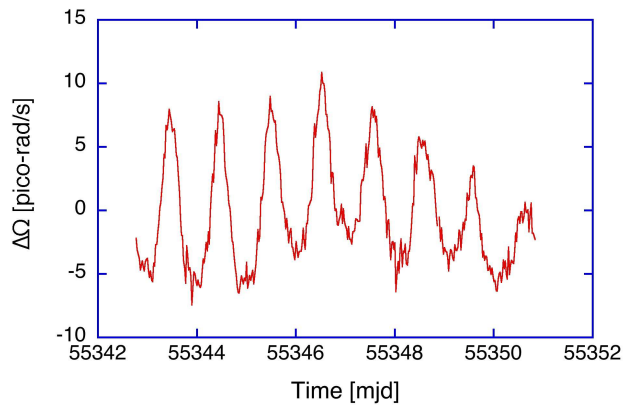}
\caption{\it Approximately eight days of raw G data taken with 30 minutes of integration time. One can clearly see the contributions from diurnal polar motion, solid Earth tides and local tilt.}
\label{G-raw}
\end{figure}

A typical eight day long measurement sequence of rotation rate data from the G ring laser is shown in fig.~\ref{G-raw}.
In order to demonstrate the obtained sensor sensitivity we have subtracted the mean Earth rotation
rate from the gyroscope data.
The y-axis gives the measured variation of the rate of rotation, while the x-axis shows the time expressed in the
form of the modified julian date. Each data point was taken by integrating over 30 minutes of measurement data.
There are several distinct signal contributions in the data, which come from known geophysical
effects. The most prominent signal is caused by diurnal polar motion \citep{polmot}. The polar motion
data is superimposed by a tilt signal caused by the semi-diurnal tides of the solid Earth,
distorting the otherwise sinusoidal diurnal frequencies slightly. At the
Geodetic Observatory in Wettzell the solid Earth tides can be as large as 40 nano-rad in amplitude.
Less evident in fig.~\ref{G-raw} are the effects from local tilt. These signals are non-periodic and
usually change slowly over the run of several days. High resolution tiltmeters inside the pressure stabilizing
vessel of the G ring laser keep track of these local effects and the data is corrected for gravitational
attraction (atmosphere, sun and moon) and corresponding changes in the gravitational potential \cite{polmot}.
Local tilts change most prominently
after abundant rainfall, indicating hydrological interactions with the rock and soil
beneath the ring laser monument. The large seasonal temperature effect on the G ring laser as well
as the substantial local tilt signals and the rather high ambient noise level of our near soil surface
structures give reasonable hope of much better performances of a ring laser installation in a deep
underground laboratory such as the Gran Sasso laboratory of INFN (Istituto Nazionale di Fisica Nucleare).

\begin{figure}[ht]
\centering
\includegraphics[height=5.4cm]{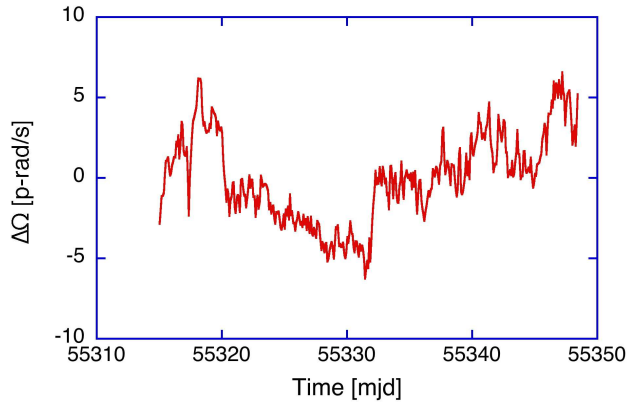}
\caption{\it The rotation rate of the Earth measured with the G ring laser as a function of time. Averaging over 2 hours was applied to a corrected dataset, where all known geophysical signals have been removed.}
\label{avg2h}
\end{figure}

For the detection of fundamental physics signals one has to remove all known perturbation signals of the Earth from the
ring laser time-series. Furthermore we have applied 2 hours of averaging of the data in order
to reduce the effect from short period perturbations. Figure~\ref{avg2h} shows an example.
In order to reduce the local orientation uncertainties, which remain after local tilts measured with the
high resolution tiltmeters have been removed,
averaging as indicated above was applied to a series of 30 days of data
collection, including the period shown in fig.~\ref{G-raw}. It can be expected that a similar data set
from the GranSasso laboratory would become substantially smoother, since most of the
perturbations, caused by ambient atmosphere - topsoil interaction still contained in the data
of fig.~\ref{avg2h} would no longer exist in the deep underground facility. Changing hydro-logic fluxes
presumably causing small local rotation, temperature variations, atmospheric pressure and wind loading
are among the sources for the systematic signatures in the residual data.

Despite the fact that large ring lasers are very
stable platforms and with the provision of tight feedback systems to stabilize the scale factor (cold cavity, as well
as the active cavity), currently ring laser gyroscope are not able
to determine the DC part of the Earth rotation rate with a sensitivity compatible with the requirements
for detection of the Lense-Thirring effect. While the contribution of the varying Earth rotation itself
presumably can be removed with sufficient accuracy from the C04 series of VLBI measurements, there
remains the problem of determining the actual null-shift offsets from the laser functions in the
ring laser gyroscope. Since the gravito-magnetic effect is small and constant, a good discrimination
against laser biases, such as for example `Fresnel drag' inside the laser cavity must be achieved.
Therefore it will be advantageous to add one or several ring laser cavities in addition to the triad
structure for sufficient redundancy. We also intend to operate at least the G ring laser structure
in parallel to the here proposed structure in order discriminate local perturbation signals from
regional and global ones. A second large ring laser located at the Cashmere facility in Christchurch,
New Zealand will be used in the data analysis process, provided it can be run with sufficient resolution
and stability.

\section{\label{sec:e}Basic Design features of a future ring laser instrument}

\begin{figure}[ht]
\includegraphics[width=271pt,angle=270]{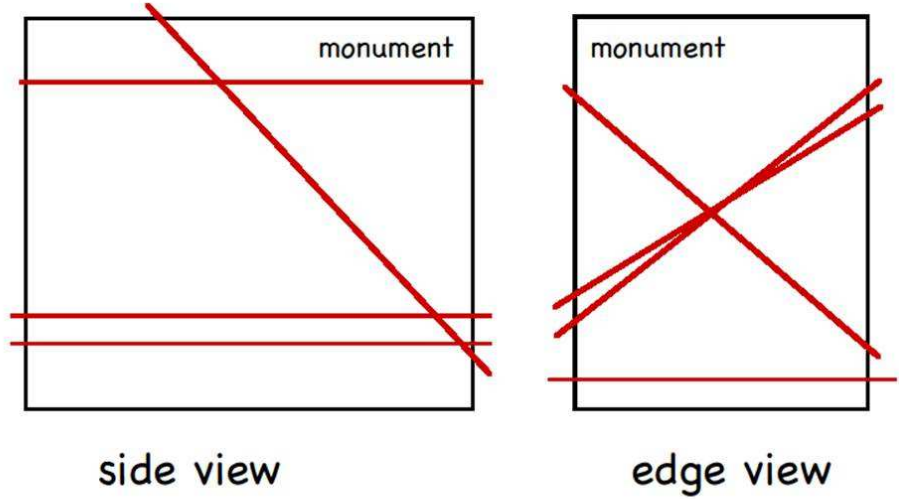}
\caption{Sketch of the experimental apparatus, the two close gyrolasers (twins) are sketched as well,
oriented near the zero Sagnac point. A gyrolaser oriented outside the meridian plane is as well
sketched, important not for the gm measurement, but to keep under control the overall motion of
the monument.}
\label{Monument}
\end{figure}

\begin{figure}[ht]
\includegraphics[width=271pt]{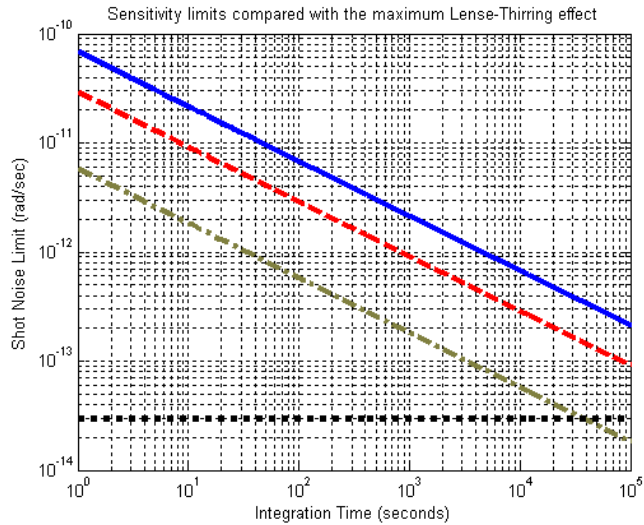}
\caption{Sensitivity of a shot noise limited gyrolaser increasing the integration time.
Blue-continuous refers to G in Wettzell (output power $20nW$), red-dotted to G with $500\mu W$
output power, green-dash-dotted is a $24$ $m$ perimeter with $500 nW$ output power.    The
horizontal black-dotted line is the maximum gm signal for a single gyrolaser}
\label{shot}
\end{figure}

We outline our first experimental design. Primarily this represents an attempt to obtain
the required accuracy in the angles by using pairs of
gyroscopes (twins), located astride the relevant positions and exploiting the
possibility to reduce (by subtraction) the big slowly changing
contributions while enhancing the weak terms whose sign changes on the two
sides of the orientation which provides zero Sagnac.
   The experimental set-up is composed of a central rigid heavy body, called a monument, to whom all
   the gyrolasers are rigidly attached,    fig. \ref{Monument} shows the set up described here.
The monument itself plays an important role in this experiment.
In order to provide a measurement base, we propose to use a massive block of concrete as a stable
central reference. Around this concrete foundation, we arrange at least 3 pairs of stainless steel
ring laser structures in the form of a square (side length $\approx 6m $), which are
firmly attached to the central monument block. Each of these ring
laser pairs has to be orientated such that a sufficient amount of the Earth rotation vector is
projected on each ring laser orientation, so that all three linearly independent directions of space
are covered. Active perimeter stabilization derived from evaluating the optical frequency of each
ring laser component is used to preserve the scale factor of each ring laser to better than 1 $nm$.  Several of such ring lasers have to be arranged and
metrologically inter-connected such, that the modulus of the instantaneous Earth rotation
vector can be extracted from the respective projections on all ring laser components. The metrology
system, constructed from several optical interferometers has to allow a proper relative orientation of the
ring at the level of better than 1 nano-radian.
A $3$ gyros system is the minimum to reconstruct the Earth angular velocity vector, but the more
gyros the better, since the noise grows with the square root of the number of independent gyros.
In the scheme analysed here we consider the construction of twin gyroscopes, i.e. two gyros supported
by a common frame, with small relative angle, as sketched in fig. \ref{Monument}. The lasers are
arranged in a way to have the vector $\hat{n}$ in the meridian plane, two sides parallel and the
other two crossing in the middle with a relative angle $\xi$, small but viable
(for instance in the following examples $\xi=1^0$).
Each pair of ring lasers can be constructed from a single mechanical frame,
which contains two mirrors per mirror holder rigidly positioned with respect to each other, sharing
the same supply of laser gas. The entire structure is as symmetrical as possible and cavity length
stirring affects both lasing contours at the same time. In this way it will be possible to use them
as a nulling interferometer and any variation of the beat note with time indicates instrumental
effects.
The relative angle of the twins can be very well controlled, and this configuration can be used to
study noise sources coming from the lasers. Knowing with $nrad$ accuracy the absolute position of
the whole system the Sagnac and Geodetic term can be subtracted and the results, which contain the
gravito-magnetic term only, can be subtracted or summed in order to have a very small number,
proportional to the derivative of gm, in practice a null measurement, and about twice the gm term,
since the two are aligned following the maximum absolute value of the curve, see fig \ref{fig2}. The
null measurement is an important tool to investigate the presence of systematics in the whole apparatus.
A system with 4 twins, for example $1^0$ one from the other, rigidly attached, improves the measurement
by a factor $4$ for the signal. Moreover with 4 independent measurement the noise (shot noise) will
be increased by a factor $2$. So, a $4$ twins system improves the measurement in total by a factor $2$.
Shot noise gives the limit in sensitivity of gyrolaser \cite{geoff}. In fig. \ref{shot} the sensitivity
in function of the integration time is given.
 Fig. \ref{shot} shows that with 27 hours of integration a single large gyro would measure gm
 at $67\%$, but if the gyros are $4$, symmetrically oriented with respect to the zero Sagnac
 point, the measurement will be at two times better. To go down to $10\%$ the integration time
 should be approximately $12$ days, while $1\%$ requires $3.2$ years. A single device would get
 the same sensitivity with more than 12 years of data taking.\\
Gyrolaser technology is approaching the sensitivity to make a direct measurement of the Earth
gravitomagnetic effect. The above described set-up is in principle very powerful, but it requires
to know with a high level of accuracy the relative position of the gyrolasers and the axis of the
Earth, which is not an easy task, especially for an underground experiment. It is worth noticing
that the LT precession can be measured comparing the modulus of the Earth rotation estimated by
the multi-gyros system and VLBI, the modulus is necessary in order to be independent of the
reference frame; in this way it is possible to reduce the requirement on the absolute orientation
of the gyroscope with respect to the instantaneous axis of rotation of the Earth. At the moment
we are investigating experimental solutions were the absolute orientation of the gyros is not
important, and only the relative positions of the gyros matter. Further study is necessary in
order to complete the whole design, which we expect to be completed in $2011$; in any case, the
final conclusions remains the same, a set of several gyros, $6$ $m$ side rigidly attached one
to the other (the more the better), is necessary in order to obtain $1\%$ sensitivity integrating
the signal for several years, the fact that the system is composed of more than $3$ independent
instruments gives the possibility to cross-check the behaviour of each instrument, and provides
linear combinations of the measurements in which the gm effect is cancelled, important check for
very low frequency drifts and any kind of $1/f$ noise.

\section{\label{acknow} Acknowledgements}

We have to acknowledge Massimo Cerdonio, for his stimulating discussions and encouragement and Maria Allegrini for her continuous and firm scientific support.

%
%
%
\end{document}